\documentclass[showpacs,preprintnumbers,preprint,aps,amsmath,amssymb]{revtex4-1}
\usepackage{color}
\usepackage{graphics}
\usepackage{amsmath,amssymb,epsfig}
\newcommand{\mathsym}[1]{{}}

\def\10{$SO(10)$}

\newcommand{\ba}{\begin{array}}
\newcommand{\ea}{\end{array}}
\newcommand{\be}{\begin{equation}}
\newcommand{\ee}{\end{equation}}
\newcommand{\beqa}{\begin{eqnarray}}
\newcommand{\eeqa}{\end{eqnarray}}
\def\321{$SU(3)\times SU(2)\times U(1)$}

\def\vev#1{\left\langle #1\right\rangle}%
\newcommand{\dms}  {\Delta m^2_{sol}}
\newcommand{\Dma}  {\Delta m^2_{atm}}

\begin{document}
\vspace*{1cm}
\title{Fermion Masses in $SO(10)$ Models}
\bigskip
\author{Anjan S. Joshipura\footnote{anjan@prl.res.in} 
and Ketan M. Patel\footnote{kmpatel@prl.res.in}} 
\affiliation{Physical Research Laboratory, Navarangpura, Ahmedabad-380 009,
India. \vskip 1.0truecm}

\begin{abstract}
\vskip 0.5 truecm 
We examine many $SO(10)$ models for their viability or otherwise in explaining all the fermion
masses and mixing angles. This study is carried out for both supersymmetric and non-supersymmetric
models and with minimal ($10+\overline{126}$) and non-minimal ($10+\overline{126}+120$) Higgs
content. Extensive numerical fits to fermion masses and mixing are carried out
in each case assuming dominance of type-II or type-I seesaw mechanism. Required
scale of the $B-L$ breaking is identified in each case. In supersymmetric case,
several sets of data at the GUT scale with or without
inclusion of finite supersymmetric corrections are used. All models studied
provide quite good fits if the type-I seesaw mechanism dominates while many fail
if the type-II seesaw dominates. This can be traced to the absence of the
$b$-$\tau$ unification at the GUT scale in these models. The minimal
non-supersymmetric model with type-I seesaw dominance gives excellent fits. In
the presence of a $45_H$ and an intermediate scale, the model can also account 
for the gauge coupling unification making it potentially interesting  model for
the complete unification. Structure of the Yukawa coupling matrices obtained
numerically in this specific case is shown to follow from a very simple $U(1)$
symmetry and a Froggatt-Nielsen singlet.
\end{abstract}


\maketitle

\section{Introduction}

Grand unified theories (GUTs) which unify strong and electroweak interactions also provide a
constrained and unified description of the fermion masses and mixing angles. This is particularly
true in case of theories based on the $SO(10)$ group \cite{early}. All the known fermions plus the
right handed (RH) neutrino of a given generation are unified into a single spinorial 16-dimensional
representation of the group. As a consequence, in a renormalizable theories of this type only three
Yukawa coupling matrices $Y_{10}$, $Y_{\overline{126}}$, $Y_{120}$ and relative strengths between
them  determine six physical mass matrices $M_f$ with $f=u,d,l$ denoting quarks and the charged
leptons, $f=D$ corresponding to the Dirac mass matrix for neutrinos and $f=L,R$ denoting the
corresponding Majorana mass matrices for the left handed and right handed neutrinos respectively.
The labels $10,\overline{126},120$ correspond to three possible Higgs representations contained in
the product $\overline{16}\times \overline{16}$. The presence of all these Higgs fields is not
necessary and more economical and allowed possibility is to choose only two of them namely, $10$
and $\overline{126}$. The consistent $SO(10)$ breaking needs additional Higgs representations -
$210$ or $210+54$ - in case of supersymmetric theories \cite{moh54} and $45$ in
case of the non-supersymmetric $SO(10)$ model.\\

The minimal supersymmetric $SO(10)$ model with Higgs fields transforming as $10,
126+\overline{126}, 210$ has limited numbers of free parameters and is explored
in all its details \cite{btau, minimal1, minimal2, bertolini1, minimal3}.
Detailed analysis of fermion spectrum is also presented in case of
the non-minimal supersymmetric model containing an additional Higgs field in the
$120$ representation of $SO(10)$ \cite{nonminimal1, mutau, grimus1, grimus2,
bertolini120, mohapatra120, charan120, altarelli}. Similar analysis in case of
non-supersymmetric $SO(10)$ model is not done and the main purpose of the
present paper is to provide such an analysis although we also give
a comprehensive discussion of various numerical fits in the supersymmetric
models.\\

The non-supersymmetric GUTs do not have built-in explanation of the  gauge hierarchy problem but
they do share several nice  features of the supersymmetric models and avoid some of the problems
associated with the  latter. They  allow gauge coupling unification \cite{bertolini3} and can also
provide dark matter candidate in the form of axion. In fact, a minimal
non-supersymmetric \10  model has been revived  in recent
studies \cite{bertolini3, bertolini4}. Earlier discussion of gauge 
coupling unification in such models is given in \cite{deshpande}. The minimal
model is more economical then the corresponding supersymmetric case as far the
choice of Higgs representation is concerned and uses only three sets of Higgs
fields namely $10_H,\overline{126}_H$ and $45_H$. This choice is argued to lead
to successful $SO(10)$ model on two counts:

\noindent (A) Two or three steps breaking of $SO(10)$ to the standard model (SM)
is possible through the vacuum expectation value (vev) of $45_H$ and
$\overline{126}_H$. An intermediate scale  $\sim 10^{11}$ GeV allows the gauge
coupling unification. This is shown through a detailed analysis using
two loop renormalization group equations \cite{bertolini3}. The presence of 
intermediate scales is also welcome from the point of view of explaining
neutrino masses. This is unlike the minimal supersymmetric model where an
intermediate scale required for neutrino masses spoils the gauge coupling
unification.

\noindent (B) The minimal model is argued \cite{bertolini4}  to be complete to the extent  that the
required pattern of the Higgs vacuum expectation values for the gauge symmetry breaking and gauge
coupling unification  can emerge from the minimization of the 1-loop corrected
Higgs potential.\\

The fermion masses in the minimal supersymmetric or non-supersymmetric models arise from the
following terms
\be
\label{minimalW}
16_F(Y_{10} 10_H+Y_{\overline{126}}\overline{126}_H)16_F ~\ee
The above terms represent a part of the superpotential in the supersymmetric case. In the
non-supersymmetric version, one would need additional Peccei-Quinn like symmetry and the above terms
would then represent the most general allowed fermion mass terms in the model. Two important
features of eq.(\ref{minimalW})  are the following.
\begin{enumerate}
 \item If the contribution of $Y_{10}$ to the masses of the third generation dominates then one gets the $b$-$\tau$
unification 
\be \label{btau}
Y_b=Y_\tau ~ \ee
This can lead to large atmospheric mixing if neutrinos obtain their masses from the type-II seesaw 
mechanism \cite{btau}.
\item Non-leading contribution to the second generation masses coming from $\overline{126}_H$ imply
a relation
\be \label{mus}
m_\mu=3 m_s \ee
between the muon and the strange quark masses.
\end{enumerate}
Both these relations are regarded as successful classical predictions of GUTs like $SO(10)$
\cite{early} or $SU(5)$ a la Georgi-Jarlskog \cite{gj}. They are supposed to hold at the grand
unification scale $M_X$. The available experimental information does not quite agree with these
generic predictions at the quantitative level. Extrapolation of the quark masses
at the GUT scale in the non-supersymmetric theories do not show $b$-$\tau$
unification. In supersymmetric theories, the evolved values of the Yukawa
couplings depend on $\tan\beta$ and it is found that eq.(\ref{btau})
holds only at some special values of $\tan\beta$. Moreover, the presence of
supersymmetry breaking around weak scale introduces additional finite
$\tan\beta$ and sparticle mass-dependent corrections which need to be included
in extrapolation. Eq.(\ref{mus}) also gets violated in a large parameter
space at $M_X$ in these theories. Several existing analysis of fermion masses in
supersymmetric models \cite{minimal1, minimal2, bertolini1, mutau, grimus1,
grimus2, altarelli} are based on simple and somewhat old extrapolation of
fermion masses presented in \cite{parida} which does not include finite
threshold corrections. As far as the non-supersymmetric theories are concerned
there has not been any exhaustive confrontation of the extrapolated values
\cite{xing} of fermion masses and mixing with the simple $SO(10)$ based models.
Motivated by this, we address three main issues in this paper. (1) We update the
existing analysis of fermion spectrum in various supersymmetric models
using the recent extrapolation of fermion masses and mixing at the GUT scale
\cite{ross1,antusch}. (2) We provide new fitting of the fermion spectrum in
various non-supersymmetric models and (3) try to understand the fitted
structure of fermion mass matrices in terms of simple patterns. This provides
hint into possible flavour structure of fermion spectrum and possible origin of
large leptonic mixing angles.\\

We begin by giving an overview of the existing analysis of fermion masses and mixing.
Then we discuss our fits in supersymmetric models. Here we consider both the
minimal and the non-minimal models. Similar analysis is then carried out in
Section IV in  the non-supersymmetric case using the input from \cite{xing}. The
analysis in these two sections demonstrates the viability or otherwise of
various $SO(10)$ models from the point of obtaining correct fermion spectrum. In
addition, it leads to very interesting structure for fermion mass matrices which
can be a starting
point to uncover the underlying flavour symmetries. We identify such structures in Section V and
present a summary of all the new results of our analysis in Section VI.\\

\section{Overview}
Masses of fermions belonging to $16$-dimensional spinorial representation of \10 arise from the
renormalizable couplings with Higgs fields belonging to $\overline{16}\times\overline{16} =
10+\overline{126}+120$ representations
\be
\label{yukawa}
16_F(Y_{10}10_H+Y_{\overline{126}}\overline{126}_H+Y_{120}120_H)16_F\ee
in a self explanatory notation. The above couplings represent terms of the superpotential in case of
supersymmetric models and Yukawa interactions in case of non-supersymmetric models with
Peccei-Quinn symmetry, see latter. $(Y_{120}),Y_{10},Y_{\overline{126}}$ are complex
(anti)symmetric matrices in
generation space. The Higgs fields $10_H+\overline{126}_H+120_H$ are forced to be complex in case of
the supersymmetric theories. The $10_H$ and $120_H$ representations of Higgs fields can be real in
non-supersymmetric  models but we shall take them to be complex in this case also to allow for the
PQ symmetry. Throughout this paper, we shall restrict to the renormalizable models. There exists
detailed analysis \cite{nr} of fermion masses in non-renormalizable models as well. The $10_H,
\overline{126}_H, 120_H$ respectively contain $1,1$ and $2$ up-like and equal number of down like
Higgs doublets. It is assumed that only one linear combination of each remains light and acquires
vacuum expectation value. This results in the following fermionic mass matrices \cite{mutau,
grimus1, grimus2}:
\beqa \label{genmass}
M_d&=& H+F+i G~,\nonumber \\
M_u&=&r (H+s F+i t_u~ G~), \nonumber\\
M_l&=& H-3 F+~it_l~ G~,\nonumber\\
M_D&=&r (H-3s F+i t_D~ G~),\nonumber\\
M_L&=& r_L F~,\nonumber\\
M_R&=& r_R^{-1} F.\eeqa
where ($G$)  $H$, $F$ are complex  (anti)symmetric matrices. $r,s,t_l,t_u,t_D,r_L,r_R$ are
dimensionless complex  parameters of which $r,r_L,r_R$ can be chosen real without lose of
generality. The effective neutrino mass matrix for three light neutrinos resulting after the seesaw
mechanism can be written as
\be \label{mnu}
{\cal M}_\nu=r_LF-r_RM_DF^{-1}M_D^T\equiv {\cal M}_\nu^{II}+{\cal M}_\nu^{I} ~.\ee
Eqs.(\ref{genmass},\ref{mnu}) describe the most general mass matrices in any renormalizable 
$SO(10)$ models and contain a large number  of parameters to be of use. Therefore various special
cases are considered in the literature and we summarize them below.\\

\subsection{Minimal Supersymmetric Model} 
This model is characterized by the absence of $120_H$ and hence $G$ in eq.(\ref{genmass}). $H$ can
be diagonalized with real and positive eigenvalues by rotating the original $16$-plets in the
generation space. Hence all the mass matrices are determined by 19 real parameters if only type-II
or type-I seesaw dominates. These parameters are determined using 18 observable
quantities. In spite of the number of observables being less than the
parameters, not all observables can be fitted with required precision due to
non-linear nature of eq.(\ref{genmass}). Eqs.(\ref{genmass}) (with $G=0$)
are fitted to the observed fermion parameters in various papers \cite{minimal1,
minimal2, bertolini1}. The most general minimization is performed by Bertolini {\it et al}
\cite{minimal1} allowing for arbitrary combination of both the type-I and II seesaw contributions
to neutrino masses. The input values for quark and the charged lepton masses used in this analysis
is taken from \cite{parida} and correspond to $\tan\beta=10$. The best fits are obtained in a mixed
scenario, type-I gives slightly worse and type-II  scenario is unable to reproduce all the
observables within $1\sigma$. If type-II seesaw dominates then one needs $b$-$\tau$ unification at
the GUT scale in order to reproduce large atmospheric mixing angle. In contrast, the extrapolated
values used in the analysis do not show complete $b$-$\tau$ unification. This results in a poor fit
to the atmospheric mixing angle at the minimum. Threshold effects can play important role in
achieving the $b$-$\tau$ unification and improves the fit to fermion masses 
compared to analysis in \cite{bertolini1} as we shall see.\\

\subsection{Non-minimal supersymmetric model} 
The other case extensively discussed in the literature corresponds to adding a 120-plet of Higgs to
the minimal model. Fermion masses in models in this category have been analyzed either assuming
type-I \cite{grimus1, grimus2, mutau, charan120} or type-II \cite{mutau, nonminimal1, altarelli,
mohapatra120, bertolini120} seesaw dominance. In this case, the most general model assuming type-II
(type-I) dominance has 29 (31) independent parameters after rotating to basis with a real and
diagonal diagonal $H$. One needs to make additional assumptions in order to reduce the parameter
space. Considerable reduction in number of parameters is achieved assuming parity symmetry
\cite{mohapatra120} or equivalently spontaneous CP violation \cite{grimus2}.
This leads to Hermitian Dirac mass matrices. In our notation, this corresponds
to taking all the parameters in eq.(\ref{genmass}) to be real, see \cite{mutau,
grimus2} for details.  Such a model has only 17 parameters in case of the
type-II dominance, two less than in case of the minimal model without $G$
but with arbitrary complex parameters. Number of parameters can be further
reduced by imposing additional discrete symmetry; $Z_2$ \cite{grimus1} or
$\mu$-$\tau$ \cite{mutau} are considered in this context. In spite of the
reduction in number of parameters the allowed fermionic structure is
analytically argued \cite{bertolini120, mohapatra120}  to help in reducing
tension in obtaining correct CP violating phase or fitting the first generation
masses.\\

Numerical fits depend on whether type-II or type-I seesaw mechanism is used. Comparison of various
models in case of the type-II seesaw dominance is made in \cite{altarelli}. All the models in this
category give a very good fit to data with a significantly lower $\chi^2$ than in case of the
minimal model. The assumption of the type-I dominance leads to better fits compared to the type-II
case. Moreover, unlike the type-II dominance, one does not need  intermediate scale \cite{mutau,
grimus1, grimus2, charan120} for reproducing the correct neutrino mass scale. This is a welcome
feature from the point of view of obtaining the gauge coupling unification. All these works are
based on the use of quark masses derived  in \cite{parida} at $\tan\beta=10$. We shall re-examine
the non-minimal model with a different set of input which include the finite threshold
corrections.\\

\subsection{Non-supersymmetric \10 models}
One common feature of all  fits with   type-II seesaw dominated scenarios is the
need for an intermediate scale $M_I\sim 10^{12}-10^{14}$ GeV. This spoils the
gauge coupling unification in
supersymmetric theories. In contrast, an intermediate scale in
non-supersymmetric  framework helps
in achieving the gauge coupling unification. But unlike the supersymmetric case, the
non-supersymmetric models do not show the $b$-$\tau$ unification and thus type-II models in this
category do not immediately explain the large atmospheric neutrino mixing angles. Viability of this
scenario can be checked through detailed numerical fits. Unlike the SUSY case, there is no
systematic and complete three generation analysis of fermion masses within non-supersymmetric
models. Various issues involved are summarized in a recent paper \cite{goran} which contains
analytic discussion of the simplified two generation case.\\

The most economical possibility for fermion masses and mixing in non-supersymmetric model would be
to choose a real $10_H$ or $120_H$ and $\overline{126}_H$ multiplets of Higgs fields. The latter is
required for neutrino mass generation but by itself, it cannot generate fermion mixing. Thus
additional $10_H$ or $120_H$ field is also needed. It is argued \cite{goran} that a
$\overline{126}_H$ and a real $10_H$ cannot fit even two generation case. Thus one needs a complex
$10_H$. Since both the real and the imaginary parts of $10_H$ can independently couple to fermions,
this would mean additional Yukawa couplings. This can be avoided by assigning a Peccei-Quinn (PQ)
charge to $10_H$. Consider the following general definition of the PQ symmetry:
\be \label{pq}
\ba{cc}
16_F\rightarrow e^{i\alpha} 16_F; &\overline{126}_H\rightarrow e^{-2i\alpha} \overline{126}_H\\
10_H\rightarrow e^{-2i\alpha} 10_H; &120_H\rightarrow e^{-2 i\alpha} 120_H\\
\ea ~.\ee
The most general Yukawa couplings allowed by this symmetry once again reduce to eq.(\ref{genmass}).
Thus formally both supersymmetric and non-supersymmetric cases look alike. But there is an important
difference. The renormalization group running of the Yukawa couplings is different in these two
cases. Moreover the non-supersymmetric case has intermediate scales. Thus input values and
consequently the resulting fits would be quite different in these two cases.\\

\section{Fermion Masses in Supersymmetric Theories: Numerical Analysis}
In this section, we present the numerical analysis of fermion masses and mixing in different
supersymmetric cases. We use the data in Table(\ref{susyinput}) and define the following $\chi^2$ function
\be \label{chisq}
\chi^2=\sum_i \left(\frac{P_i-O_i}{\sigma_i}\right)^2 ~,\ee
where the sum  $i=1,..,14$ runs over seven mass ratios and four quark mixing parameters (given in
Table(\ref{susyinput})), ratio of the solar to atmospheric mass squared differences and the solar 
($\theta_{12}^l$) plus the atmospheric ($\theta_{23}^l$) mixing angles \cite{fnt}. For the latter we
use the values given in \cite{numass}. These data are fitted by numerically minimizing the function
$\chi^2$. We assume $\Delta m^2_{atm}$ to be positive corresponding to the normal neutrino mass
hierarchy. We also impose the $3\sigma$ upper bound on $\theta_{13}$ while minimizing the $\chi^2$.
$P_i$ denote the theoretical values of observables determined by the input
expression, eq.(\ref{genmass}) and $O_i$ are the experimental values extrapolated to the GUT scale.
$\sigma_i$ denote the errors in $O_i$.\\

\subsection{Minimal Supersymmetric model}
Our input values of  the quark masses  and mixing angles at the GUT scale are based on the analysis 
in \cite{ross1}. This uses more precise values of the $b$ and $t$ quark masses and the CKM
parameters. More importantly, finite threshold corrections induced by sparticles are included in
this analysis. Analysis in \cite{ross1} proceeds in two steps. First, the quark masses and mixing
angles are determined by fitting the available low energy data and evolving them to the
supersymmetry braking scale $M_S$. In the second step, finite sparticle induced corrections are
included and then evolution is performed up to the GUT scale $M_X$. These corrections are expressed
in terms of phenomenological parameters $\gamma_{d,b,u,t}$ defined below. We reproduce their table
of values so obtained  as Table(\ref{susyinput}) for convenience of the reader. 
\begin{table}[ht]
\centerline{
\begin{tabular}{|c|c|c|c|c||c|c|}
\hline
 $ $ & {\bf A} & {\bf B} & {\bf C} &  {\bf D} & {\bf C1} & {\bf C2} \cr
\hline
 $\tan \beta $ & $1.3$ & $10$ & $38$ &  $50$ & $38$ & $38$ \cr
 $\gamma_b$ & $0$ & $0$ & $ 0$ &  $0$ & $-0.22$ & $+0.22$ \cr
 $\gamma_d$ & $0$ & $0$ & $ 0$ &  $0$ & $-0.21$ & $+0.21$ \cr
 $\gamma_t$ & $0$ & $0$ & $ 0$ &  $0$ & $0$ &  $-0.44$ \cr
 $y^t(M_X)$ & $6^{+1}_{-5}$ & $0.48(2)$ & $0.49(2)$ & $0.51(3)$               	 & $0.51(2)$ &
$0.51(2)$ \cr
  $y^b(M_X)$ & $0.0113^{+0.0002}_{-0.01}$ & $0.051(2)$ & $0.23(1)$ & $0.37(2)$ 	 & $0.34(3)$
& $0.34(3)$ \cr
  $y^\tau(M_X)$ & $0.0114(3)$ &  $0.070(3)$  & $0.32(2)$ & $0.51(4)$          	 & $0.34(2)$ &
$0.34(2)$   \cr
 \hline
Observables & \multicolumn{6}{|c|}{GUT scale values
 with propagated uncertainty }  \cr
 \hline
  $(m_u/m_c)$ & $0.0027(6)$ & $0.0027(6)$ & $0.0027(6)$ & $0.0027(6)$     	 & $0.0026(6)$ &
$0.0026(6)$ \cr
  $(m_d/m_s)$ & $0.051(7)$ & $0.051(7)$ & $0.051(7)$ & $0.051(7)$        	 & $0.051(7)$  &
$0.051(7)$   \cr
  $(m_e/m_\mu)$ & $0.0048(2)$ & $0.0048(2)$ & $0.0048(2)$ & $0.0048(2)$  	 & $0.0048(2)$ &
$0.0048(2)$ \cr
  $(m_c/m_t)$ & $0.0009^{+0.001}_{-0.00006}$  & $0.0025(2)$ & $0.0024(2)$ & $0.0023(2)$  	&
$0.0023(2)$ & $0.0023(2)$ \cr
  $(m_s/m_b)$  & $0.014(4)$ & $0.019(2)$ & $0.017(2)$ & $0.016(2)$       	 & $0.018(2)$ &
$0.010(2)$ \cr
  $(m_\mu / m_\tau)$ & $0.059(2)$  & $0.059(2)$ & $0.054(2)$ & $0.050(2)$      	 &
$0.054(2)$ & $0.054(2)$ \cr
  $(m_b/m_\tau)$
     & $1.00^{+0.04}_{-0.4}$    & $0.73(3)$     & $0.73(3)$     & $0.73(4)$        & $1.00(4)$
& $1.00(4)$ \cr
  $\sin \theta_{12}^q$& $0.227(1)$ & $0.227(1)$ & $0.227(1)$ & $0.227(1)$          	 &
$0.227(1)$ & $0.227(1)$ \cr
  $\sin \theta_{23}^q$ & $0.0289^{+0.0179}_{-0.00073}$ & $0.0400(14)$ & $0.0386(14)$
&
$0.0371(13)$ & $0.0376(19)$ & $0.0237(18)$ \cr
  $\sin \theta_{13}^q$ & $0.0026^{+0.0022}_{-0.00045}$ & $0.0036(7)$ & $0.0035(7)$
& $0.0033(7)$   & $0.0034(7)$& $0.0021(5)$  \cr
  $\delta_{CKM}[^{\circ}]$ & $56.31\pm10.24$ & $56.31\pm10.24$ & $56.31\pm10.22$ & $56.31\pm10.22$  
        & $56.31\pm10.27$ & $56.31\pm10.25$ \cr
  \hline
 \end{tabular}}
\caption{The input values of various observables of quark sector and charged lepton masses obtained
at GUT-scale $M_X$ for various values of $\tan \protect\beta $ and threshold corrections
$\protect\gamma_{t,b,d}$ assuming an effective SUSY scale $M_{S}=500$ GeV (see \cite{ross1} for
details).}
\label{susyinput}
\end{table}

Column (A)-(D) show the evolved values of quark mass ratios and mixing angles  
in the absence of
threshold corrections for various values of $\tan\beta$.  One clearly sees the absence of the
$b$-$\tau$ unification at the GUT scale except for the low value of $\tan\beta$. This changes with
the inclusion of threshold corrections. These corrections are parameterized by $\gamma_{d,u,b,t}$
which are defined in the following manner. The down quark mass matrix is determine  by the term
$QY_dd^cH_d$ in the minimal supersymmetric standard model. The corresponding term $QY_d'd^cH_u^*$
involving the second doublet $H_u^*$ is not allowed in the superpotential by SUSY but it can be
radiatively generated after the SUSY breaking. Since $\tan\beta\equiv\frac{\vev
{H_u^0}}{\vev{H_d^0}}$ , such terms give significant corrections to the tree level values for large
$\tan\beta$ and should be included in evolving fermion masses and mixing from low energy scale to
the $M_X$. The corrected down quark matrix is parameterized in \cite{ross1,raby} by
$$U_L^{d\dagger}(1+\Gamma^d+V_{CKM}^\dagger\Gamma_uV_{CKM})Y^d_{\rm diag}U_R^d$$
where $U_{L,R}^d$ and $V_{CKM}$ are the (diagonal) down quark mass and the CKM matrix before the
radiative corrections. The loops involving down squark-gaugino generate the second term and the
loop with up squark-chargino generate the second term. $\Gamma_{d,u}$ are diagonal in the
approximation of taking diagonal squark masses in the basis with diagonal quarks. Assuming equality
of the first two generation squark masses, the  diagonal
elements $\Gamma_d=(\gamma_d,\gamma_d,\gamma_b)$ correct the down quark masses and 
$\Gamma_u=(\gamma_u,\gamma_u,\gamma_t)$ correct the CKM matrix in addition. The SUSY threshold
corrections are included through these parameters and their best fit values corresponding to  three
classical GUT predictions namely eq.(\ref{btau}) and eq.(\ref{mus}) and the relation
$\frac{m_d}{3 m_e}=1$ are determined. Last two columns correspond to different values of $\gamma$'s
determined this way. Comparison of column C with C1,C2 shows that threshold corrections change
significantly the $b$ quark mass as well as $\theta_{23}^q,\theta_{13}^q$. The neutrino masses and
mixing that we use are the updated low scale values \cite{numass} but the effects of the evolution
to $M_{GUT}$ on the ratio of the solar to atmospheric mass scale and on the mixing angles are known
to be small for the normal hierarchical spectrum that we obtain here.\\

We now discuss detailed fits to fermion masses and mixing based on the input values in
Table(\ref{susyinput}). We assume that either the type-I or the type-II seesaw term in the neutrino
mass matrix dominates and carry out analysis separately in each of these two cases. We can rewrite
eq(\ref{genmass}) as follows.
\beqa \label{genmassmin}
M_u&=&r  m_\tau\left( \dfrac{3+s}{4} \tilde{M}_d+\dfrac{1-s}{4} \tilde{M}_l\right) , \nonumber\\
M_D&=&r m_\tau \left(\dfrac{3(1-s)}{4} \tilde{M}_d+\dfrac{1+3s}{4} \tilde{M}_l\right),\nonumber\\
M_L&=&  \dfrac{r_L m_\tau}{4} (\tilde{M}_d-\tilde{M}_l) ,\nonumber\\
M_R&=&  \dfrac{r_R^{-1}m_\tau}{4} (\tilde{M}_d-\tilde{M}_l).\eeqa
We have chosen the basis with a diagonal $M_l$ and introduced 
$\tilde{M}_{d,l}=\frac{1}{m_\tau}M_{d,l}$. Thus
$$\tilde{M}_l=Diag.(m_e/m_{\tau},m_{\mu}/m_{\tau},1)$$
Hence all the quantities in the bracket in the above equation depend on the  known ratios of
charged lepton masses. $\tilde{M}_d$ is a complex symmetric matrix with 12 real parameters. Since we
are fitting the ratios of different mass eigenvalues and mixing angles, the parameter $r$ remains
free and it can be fixed by $m_t$. $r_L$ ($r_R$) in the case of type-II (type-I) seesaw dominance is
determined from the atmospheric mass scale. We have total 14 real parameters (12 in $\tilde{M}_d$
and complex $s$) which are fitted over 14 observables. Four unknown observables in lepton sector
($\theta_{13}^l$ and three CP violating phases) get determined at the minimum. Results of numerical
analysis carried out separately for the type-II and the type-I dominated seesaw mechanisms are shown
in Table(\ref{smint2}) and Table(\ref{smint1}) respectively. Let us comment on
the results.\\
\begin{table}[ht]
\begin{small}
\begin{math}
\begin{array}{|c||c|c|c|c||c|c|}
\hline
  & \text{\bf A} & \text{\bf B} & \text{\bf C} & \text{\bf D} & \text{\bf C1} &
\text{\bf C2} \\
\hline
 \text{Observables} & \multicolumn{6}{|c|} {\text{Pulls obtained for best fit solution}} \\
\hline
 (m_u/m_c) & -0.00668428 & 0.0276825 & 0.0259467 & 0.120767 & -0.0212532 &
0.0356043 \\
 (m_c/m_t) & 0.56521 & 0.157569 & 0.0201093 & 0.0730136 & 0.130288 & 0.320944
\\
 (m_d/m_s) & -1.21642 & -0.891034 & -0.27664 & -1.36265 & -1.04724 &
-1.57673 \\
 (m_s/m_b) & 0.112798 & 0.440678 & 0.163272 & 0.752408 & 0.884723 &
0.789053 \\
 (m_e/m_{\mu}) & 0.0590249 & -0.00627804 & 0.3944 & 0.0396087 & 0.0297987 &
0.0555931 \\
 (m_{\mu}/m_{\tau}) & 0.182548 & 0.103214 & 0.821485 & 0.0192305 & 0.26316 & 0.121145 \\
 (m_b/m_{\tau}) & 0.87282 & 2.20829 & 2.79368 & 2.34331 & 0.26656 & 0.407798 \\
 \left( \dfrac{\dms}{\Dma}\right)  & 0.256292 & 0.116314 & -0.14908 & 0.230056 & 0.0188227 &
-0.0140039 \\
 \sin  \theta _{12}^{q} & 0.0730813 & 0.0702755 & 0.0399788 & 0.105989 & 0.0779176 &
0.127757 \\
 \sin  \theta _{23}^{q} & -0.0311676 & -0.172792 & -0.471738 & -0.0960437 & -0.757038 &
-0.945821 \\
 \sin  \theta _{13}^{q} & 1.33502 & -0.0354198 & 0.494732 & 0.606606 & 0.890741 & 1.17758
\\
 \sin ^2 \theta _{12}^{l} & 0.00836789 & -0.106439 & -0.599727 & -0.27881 & -0.63356 &
-0.510182 \\
 \sin ^2 \theta _{23}^{l} & -1.53367 & -4.97038 & -4.95673 & -4.70944 & -2.56294 &
-1.84412 \\
 \delta _{\text{CKM}}[^{\circ}] & -0.345931 & -0.163765 & -0.600814 & -0.214459 & -0.650554 &
-0.75885\\
\hline
\chi^2_{min} & {\bf 6.9367} & {\bf 30.70} & {\bf 34.52} & {\bf 30.68} & {\bf 10.804} & {\bf 9.3559}
\\
\hline
\hline
 \text{Observables} & \multicolumn{6}{|c|} {\text{Corresponding Predictions at GUT scale}} \\
\hline
 \sin ^2 \theta _{13}^{l} & 0.0226508 & 0.0190847 & 0.0206716 & 0.0196974 & 0.0239619 &
0.0209208 \\
 \delta _{\text{MNS}}[^{\circ}] & 19.9399 & 18.9784 & 19.5619 & 11.92 & 358.789 & 1.78569 \\
 \alpha _1[^{\circ}] & 337.171 & 346.627 & 344.795 & 350.595 & 12.4786 & 349.711 \\
 \alpha _2[^{\circ}] & 147.364 & 151.912 & 146.886 & 161.702 & 194.023 & 168.156\\
 r_L m_{\tau}[\rm{GeV}] & 8.37\times10^{-10} & 6.0\times10^{-10} & 6.49\times10^{-10} &
6.94\times10^{-10} & 7.15\times10^{-10} & 9.1\times10^{-10}\\
\hline
\end{array}
\end{math}
\end{small}
\vspace{0.0cm}
\caption{Best fit solutions for fermion masses and  mixing obtained assuming the type-II seesaw
dominance in the minimal SUSY \10 model. Pulls of various observables and predictions obtained at the minimum are shown for six different data sets.}
\label{smint2}
\end{table}

\begin{table}[ht]
\begin{small}
\begin{math}
\begin{array}{|c||c|c|c|c||c|c|}
\hline
  & \text{\bf A} & \text{\bf B} & \text{\bf C} & \text{\bf D} & \text{\bf C1} &
\text{\bf C2} \\
\hline
 \text{Observables} & \multicolumn{6}{|c|} {\text{Pulls obtained for best fit solution}} \\
\hline
 (m_u/m_c) & 0.0486938 & -0.180782 & 0.0653101 & 0.0053847 & 0.0467579 &
-0.0119661 \\
 (m_c/m_t) & 1.22599 & 0.130589 & 0.246294 & 0.146932 & 0.297256 & 0.273346 \\
 (m_d/m_s) & -0.229546 & -0.730641 & 0.223201 & -0.748148 & -2.2904 &
-0.689684 \\
 (m_s/m_b) & -0.932536 & -0.886438 & -0.977249 & -1.05766 & 0.735548 &
0.000467775 \\
 (m_e/m_{\mu}) & 0.0340323 & 0.442759 & 0.103692 & -0.476364 & 0.0649144 &
-0.0648856 \\
 (m_{\mu}/m_{\tau}) & 0.310305 & -0.526529 & 0.881934 & 0.938701 & 0.705648 & 0.0178824
\\
 (m_b/m_{\tau}) & -0.486477 & -0.194215 & 0.0172182 & -0.34079 & 0.789868 &
-0.734937 \\
 \left( \dfrac{\dms}{\Dma}\right)  & 0.122267 & -0.10063 & -0.00563647 & -0.120429 & -0.180164 &
0.158557
\\
 \sin  \theta _{12}^{q} & 0.0432634 & 0.227948 & 0.0186715 & 0.084149 & 0.130301 &
0.0922391 \\
 \sin  \theta _{23}^{q} & -0.281221 & -0.0401177 & -0.167224 & 0.0649082 & -0.273222 &
-1.17651 \\
 \sin  \theta _{13}^{q} & 1.37864 & -0.275689 & 0.926186 & 0.559003 & 1.48675 & 0.248759
\\
 \sin ^2 \theta _{12}^{l} & -0.0528379 & -0.0598219 & -0.38133 & -0.172148 & -0.746107 &
0.0694831 \\
 \sin ^2 \theta _{23}^{l} & -1.22555 & -1.27077 & -1.43475 & 0.0548963 & -1.99485 &
-0.946001 \\
 \delta _{\text{CKM}}[^{\circ}] & -0.291137 & 0.397159 & -0.350422 & -0.755859 & -0.956628 &
-0.3197\\
\hline
\chi^2_{min} & {\bf 6.3479} & {\bf 3.7962} & {\bf 5.0715} & {\bf 3.8665}& {\bf 14.789} & {\bf
3.4746}\\
\hline
\hline
 \text{Observables} & \multicolumn{6}{|c|} {\text{Corresponding Predictions at GUT scale}} \\
\hline
 \sin ^2 \theta _{13}^{l} & 0.0223307 & 0.0194886 & 0.0218753 & 0.0186789 & 0.0253152 &
0.0205366 \\
 \delta _{\text{MNS}}[^{\circ}]& 2.41793 & 4.52493 & 6.08769 & 335.07 & 357.142 & 14.7651 \\
 \alpha _1[^{\circ}]& 347.106 & 8.42838 & 7.64991 & 28.0261 & 14.5679 & 1.13126 \\
 \alpha _2[^{\circ}]& 163.759 & 191.241 & 188.713 & 218.586 & 196.273 & 177.828\\
 r_R \left( \dfrac{m_t^2}{m_{\tau}}\right) [\rm{GeV}] & 1.77\times10^{-10} & 2.63\times10^{-10} &
2.50\times10^{-10} &
4.02\times10^{-10} & 7.3\times10^{-11} & 2.82\times10^{-10}\\
\hline
\end{array}
\end{math}
\end{small}
\vspace{0.0cm}
\caption{Best fit solutions for fermion masses and  mixing obtained assuming the type-I seesaw
dominance in the minimal SUSY \10 model. Pulls of various observables and predictions obtained at the minimum are shown for six different data sets.}
\label{smint1}
\end{table}

\begin{itemize}
\item  The best fit in the type-II case is obtained at low $\tan\beta=1.3$~. This case has
$b$-$\tau$ unification and threshold corrections are not very significant. On the other hand, cases
B, C, D with relatively large $\tan\beta$ but without inclusion of threshold correction give quite
bad fit. There is a clear correlation  between the overall fit and the presence or absence of the
$b$-$\tau$ unification in type-II models. Cases corresponding to the absence of the $b$-$\tau$
unification cannot reproduce the atmospheric mixing angle and results in relatively poor fits.
Inclusion of threshold corrections improves the fit but still $\frac{m_d}{m_s}$ and the atmospheric
mixing angle cannot be reproduced within $1\sigma$. The fit for $\tan\beta=10$ obtained here with
inputs from \cite{ross1,numass} is poor compared to the  corresponding fit presented in
\cite{bertolini1} which uses input from \cite{parida}. Compared to data in \cite{parida}, the result
from \cite{ross1} display larger deviation from the $b$-$\tau$ unification and also errors in more
recent input that we use for $\sin^2\theta_{23}^l$ are smaller. Both these features combine to give
larger pulls for the ratio $\frac{m_{b}}{m_\tau}$ and $\sin^2\theta_{23}^l$ and results in poor
fit. 
\item In contrast to the type-II case, the fits obtained in type-I case are uniformly better. Here
one does not expect correlation between the atmospheric mixing angle and $b$-$\tau$ unification.
Thus the cases B, C, D with large $\tan\beta$ also give quite good fits. Even in these cases
(except D) main contribution to $ \chi^2 $ comes from the pull in the atmospheric mixing angle.
Threshold corrections are significant for large $\tan\beta$ and specific cases C1, C2 achieve
$b$-$\tau$ unification but the overall fit worsens compared to B, C, D. Unlike in the type-II
case, the $\chi^2$ value obtained here for $\tan\beta=10$ is comparable to the corresponding value
in \cite{bertolini1}.
\item We have fixed the overall scale of neutrino mass $r_L (r_R)$ in the case of type-II (type-I)
seesaw by using the atmospheric scale as normalization. The resulting values are displayed in
Table(\ref{smint2}, \ref{smint1}). $r_L (r_R)$ arise from the vev of the components of
$\overline{126}_H$ transforming as $(3,1,-2)$ ($(1,3,-2)$) under the $SU(2)_L\times SU(2)_R\times
U(1)_{B-L}$. In particular, $\vev{(1,3,-2)}_{\overline{126}_H}$ sets the scale of the $B-L$ breaking
and is directly determined from the fits to fermion masses in the type-I scenario. From
eq.(\ref{genmassmin}),
\be \label{scale}
\vev{(1,3,-2)}_{\overline{126}_H}\approx r_R^{-1}v s_m\cos\beta~,\ee
where $s_m$ gives the mixing of the  light  $H_d$ in the doublet part of $\overline{126}_H$ and ~~~~
$v\approx 174$ GeV. $r_R$ is roughly independent of the input data set and for the value $r_R\approx
2.6 \times 10^{-10} m_\tau/m_t^2 ~~{\rm GeV}$, eq.(\ref{scale}) gives
$$ \vev{(1,3,-2)}_{\overline{126}_H}\approx 3.7 \times 10^{15} s_m\cos\beta ~~{\rm GeV}$$
Thus the $B-L$ breaking scale in the type-I seesaw can be close to the GUT scale for $s_m \cos\beta\sim {\cal O}(1)$.
It would however be significantly lower for large values of $\tan\beta$ and
would conflict with the constraints from the gauge coupling unification. The
determination of the $B-L$ breaking  scale in the type-II dominated scenario is
dependent on the details of the superpotential. Earlier
\cite{minimal2,bertolini1} analysis in the minimal model has shown
that this scale cannot easily be lifted to the GUT scale and pauses a problem
with the gauge coupling unification in the minimal scenario both for the type-I
and type-II seesaw dominance \cite{bertolini1}. Thus one does need to go beyond
the minimal model and models with $120_H$ are possible examples.\\
\end{itemize}

\subsection{Numerical Analysis: Extended model with $10+\overline{126}+120$ Higgs}
We now consider the non-minimal case obtained from eq.(\ref{genmass}) by choosing all parameters
in $H,F,G$ as well as $r,s,t_u,t_l,t_D,r_l,r_R$ real. As before, the parameters $r$ and $r_R$($r_L$)
determine the $m_t$ and overall scale of neutrino masses in type-I (type-II) seesaw dominated
scenarios. Our choice of 14 observables is the same as in the previous subsection. But they are now
determined from the more general expression with non-zero $G$. $H$ can be made diagonal without loss
of generality. The mass matrices $M_u,M_d,M_l,M_D,M_R$ and $M_L$ are expressed in terms of 16 real
parameters (3 in $H$, 6 in $F$, 3 in $G$, $s,t_l,t_u,$ and $t_D$ ) which determine 14 observables
$P_i$ defined before.
\begin{table}[ht]
\begin{small}
\begin{math}
\begin{array}{|c||c|c|c|c||c|c|}
\hline
  & \text{\bf A} & \text{\bf B} & \text{\bf C} & \text{\bf D} & \text{\bf C1} &
\text{\bf C2} \\
\hline
 \text{Observables} & \multicolumn{6}{|c|} {\text{Pulls obtained for best fit solution}} \\
\hline
 (m_u/m_c) & 0.00196316 & 0.019005 & -0.026015 & -0.00109589 & -0.00812155 &
0.0000225717 \\
 (m_c/m_t) & 0.000750815 & -0.114469 & 0.0964863 & 0.296526 & -0.0278823 &
-0.00413523 \\
 (m_d/m_s) & 0.0547314 & 0.618531 & 0.0606721 & -1.14305 & 0.0271889 &
0.00312586 \\
 (m_s/m_b) & 0.0565403 & 0.473347 & -1.30774 & 0.675173 & 0.0105556 &
0.0361755 \\
 (m_e/m_{\mu}) & 0.0114456 & -0.0155357 & 0.0482971 & -0.00371258 & 0.00167012 &
0.000128709 \\
 (m_{\mu}/m_{\tau}) & -0.00279654 & -0.66999 & 0.111235 & -0.0605957 & 0.00155096 &
-0.00249006 \\
 (m_b/m_{\tau}) & -0.171035 & -0.301056 & -0.381508 & 1.57926 & 0.0514536 &
-0.048487 \\
 \left( \dfrac{\dms}{\Dma}\right)  & 0.00833338 & -0.297416 & 0.191515 & 0.2129 & 0.00050834 &
-0.00412658
\\
 \sin  \theta _{12}^{q} & -0.0106839 & -0.0145213 & -0.0420229 & 0.0809603 & -0.00715584
& 0.0000538731 \\
 \sin  \theta _{23}^{q} & -0.00295777 & -0.058218 & 0.301593 & 0.341191 & 0.0120366 &
-0.000633901 \\
 \sin  \theta _{13}^{q} & -0.00466345 & -0.661544 & 0.381317 & -0.632744 & -0.137308 &
0.00650479 \\
 \sin ^2 \theta _{12}^{l} & 0.0106277 & -0.194399 & 0.333404 & 0.399294 & 0.00217496 &
-0.0043514 \\
 \sin ^2 \theta _{23}^{l} & -0.0198083 & 1.08433 & -0.472589 & -0.885401 & 0.0314484 &
0.00752103 \\
 \delta _{\text{CKM}}[^{\circ}] & -0.00915099 & 0.168633 & -0.520071 & 0.246618 & -0.0314877 &
-0.0382519\\
\hline
\chi^2_{min} & {\bf 0.0364} & {\bf 2.9315} & {\bf 2.7639}& {\bf 5.921} & {\bf 0.0254} & {\bf 0.0038}
\\
\hline
\hline
 \text{Observables} & \multicolumn{6}{|c|} {\text{Corresponding Predictions at GUT scale}} \\
\hline
 \sin ^2 \theta _{13}^{l} & 0.0215726 & 0.0312498 & 0.03568 & 0.0214329 & 0.0289663 &
0.0069694 \\
 \delta _{\text{MNS}}[^{\circ}] & 34.3864 & 5.21955 & 89.5 & 315.898 & 355.507 & 75.6953 \\
 \alpha _1[^{\circ}] & 6.26083 & 76.0772 & 289.921 & 80.1968 & 60.3609 & 240.526 \\
 \alpha _2[^{\circ}] & 161.011 & 253.288 & 76.0613 & 283.63 & 220.306 & 34.4702\\
 r_L m_{\tau}[\rm{GeV}] & 1.27\times10^{-9} & 9.57\times10^{-10} & 6.82\times10^{-10} &
1.56\times10^{-9} & 2.36\times10^{-9} & 3.68\times10^{-9}\\
\hline
\end{array}
\end{math}
\end{small}
\vspace{0.0cm}
\caption{Best fit solutions for fermion masses and  mixing obtained assuming the type-II seesaw
dominance in the non-minimal SUSY \10 model. Pulls of various observables and predictions obtained
at the minimum are shown for six different data sets.}
\label{sscpvt2}
\end{table}

\begin{table}[ht]
\begin{small}
\begin{math}
\begin{array}{|c||c|c|c|c||c|c|}
\hline
  & \text{\bf A} & \text{\bf B} & \text{\bf C} & \text{\bf D} & \text{\bf C1} &
\text{\bf C2} \\
\hline
 \text{Observables} & \multicolumn{6}{|c|} {\text{Pulls obtained for best fit solution}} \\
\hline
 (m_u/m_c) & -0.0151499 & 0.0262493 & -0.0019449 & -0.00461056 & 0.00542513 &
-0.0000775584 \\
 (m_c/m_t) & -0.000384519 & 0.000812518 & 0.000258262 & 0.00461629 &
-0.00304033 & 0.0049013 \\
 (m_d/m_s) & -0.0778857 & -0.0653974 & -0.0053692 & 0.0272334 & 0.00701785
& 0.00147573 \\
 (m_s/m_b) & -0.052311 & 0.0706689 & 0.0726379 & 0.0830354 & 0.0120634 &
0.0296307 \\
 (m_e/m_{\mu}) & 0.00127584 & 0.00152407 & 0.00164957 & 0.0268034 & -0.00722174 &
-0.000349497 \\
 (m_{\mu}/m_{\tau}) & -0.0553488 & -0.0188764 & -0.0212797 & -0.0282999 & -0.00866254 &
-0.00875371 \\
 (m_b/m_{\tau}) & -0.0103881 & 0.0214596 & 0.0260868 & 0.0498589 & 0.00493891 &
0.00967378 \\
 \left( \dfrac{\dms}{\Dma}\right)  & 0.0324886 & 0.00926157 & 0.00312614 & -0.00630473 & -0.00302065
&
0.000653399 \\
 \sin  \theta _{12}^{q} & 0.0159112 & -0.0140628 & -0.000195379 & 0.00791696 & -0.0171517
& -0.000184021 \\
 \sin  \theta _{23}^{q} & 0.0375281 & -0.00674466 & -0.00216987 & 0.00501282 & 0.0100126
& 0.00590551 \\
 \sin  \theta _{13}^{q} & 0.0309917 & 0.0571306 & 0.175888 & 0.0213394 & -0.131639 &
-0.00184989 \\
 \sin ^2 \theta _{12}^{l} & 0.00539037 & -0.0176765 & 0.00577816 & -0.013618 & 0.0092152
& 0.000404734 \\
 \sin ^2 \theta _{23}^{l} & 0.0332756 & 0.0143127 & 0.0125096 & 0.0200216 & 0.00356131 &
0.00026684 \\
 \delta _{\text{CKM}}[^{\circ}] & -0.0585649 & -0.00882152 & -0.0406312 & -0.0292954 & -0.0291351 &
-0.0310722\\
\hline
\chi^2_{min} & {\bf 0.0204} & {\bf 0.0150}& {\bf 0.0392} & {\bf 0.0137} & {\bf 0.0191}& {\bf 0.0011}
\\
\hline
\hline
 \text{Observables} & \multicolumn{6}{|c|} {\text{Corresponding Predictions at GUT scale}} \\
\hline
 \sin ^2 \theta _{13}^{l} & 0.0122064 & 0.0168745 & 0.0146633 & 0.0359278 & 0.0246489 &
0.030277 \\
 \delta _{\text{MNS}}[^{\circ}] & 87.6747 & 22.8731 & 330.351 & 282.035 & 272.186 & 84.0238 \\
 \alpha _1[^{\circ}] & 5.82048 & 167.229 & 192.077 & 286.062 & 352.828 & 329.804 \\
 \alpha _2[^{\circ}] & 339.846 & 331.88 & 34.5585 & 336.358 & 17.6585 & 325.182\\
 r_R \left( \dfrac{m_t^2}{m_{\tau}}\right) [\rm{GeV}] & 6.56\times10^{-15} & 1.22\times10^{-12} &
1.34\times10^{-12} &
3.03\times10^{-15} & 5.0\times10^{-14} & 1.40\times10^{-13}\\
\hline
\end{array}
\end{math}
\end{small}
\vspace{0.0cm}
\caption{Best fit solutions for fermion masses and  mixing obtained assuming the type-I seesaw
dominance in the non-minimal SUSY \10 model. Pulls of various observables and predictions obtained at the minimum are shown for six different data sets.}
\label{sscpvt1}
\end{table}

Results of numerical analysis carried out separately for the type-II and type-I dominated seesaw
mechanisms are shown in Table(\ref{sscpvt2}) and Table(\ref{sscpvt1}) respectively. The following
remarks are in order in connection with the results presented in these tables. As discovered in
earlier numerical analysis \cite{mutau,altarelli}, the introduction of the $120_H$ leads to
remarkable improvement in numerical fits in the type-II case. This mainly arises because the  near
maximality $\theta_{23}^l$ is not directly connected to the the $b$-$\tau$ unification. Thus 
the cases B, C, D which do not have the $b$-$\tau$ unification also lead to very good fits in
contrast to the minimal case. The fits in cases (A, C1, C2)  which have $b$-$\tau$ unification are
even better and all the observables are fitted almost exactly in these cases. These include the low
$\tan\beta$ inputs and cases with large $\tan\beta$ and threshold corrections. As the results of 
Table(\ref{sscpvt1}) show, the fits obtained assuming the type-I seesaw dominance are uniformly
better compared to the corresponding type-II results and show significantly improvement over the
minimal model with type-I dominance, Table(\ref{smint1}).\\

One important difference compared to the minimal case is the overall $B-L$ scale
determined from the neutrino masses. Unlike the minimal case, the values of
$r_R^{-1}$ in Table(\ref{sscpvt1}) are strongly dependent on the input data set
and in some cases are quite large although each data set appear to give very
good fit to fermion masses. For example, one obtains in case (A) from
eq.(\ref{scale}) and Table(\ref{sscpvt1}),
$$ \vev{(1,3,-2)}_{\overline{126}_H}\approx 1.5\times10^{20}s_m\cos\beta~{\rm GeV}$$
Thus reproducing neutrino masses in this case would require fine tuning $s_m\sim 10^{-4}$ if the
$B-L$ breaking scale is to be close to $M_{GUT}$. In contrast, in case C2 with $\tan\beta=38$,
Table(\ref{sscpvt1}) gives 
$$ \vev{(1,3,-2)}_{\overline{126}_H}\approx 1.8\times10^{17}s_m~{\rm GeV}$$
which is close to the GUT scale.\\

\section{Fermion masses in non-supersymmetric models: Numerical analysis}
We now turn to the discussion of various non-supersymmetric models. We shall consider three different cases.
\begin{enumerate}
\item The minimal scenario containing the Higgs representations $45_H+10_H+\overline{126}_H$.
\item Alternative model with $45_H+120_H+\overline{126}_H$ proposed and analyzed in case of two generations in \cite{goran}
 \item The non-minimal scenario with $45_H+10_H+\overline{126}_H+120_H$ with Hermitian structure.
\end{enumerate}
The case (2) is found to be unable to fit all the fermion masses and mixing
angles. The minimal case works quite well in this regards and there is no real
motivation to go to the non-minimal case as far as the fermion masses are
concerned. We have included this for completeness and find that this case works
even better than the minimal case.\\

The $SO(10)$ breaking \cite{bertolini4} and the gauge coupling unification
\cite{bertolini3} with intermediate scale has been reanalyzed recently following
earlier works \cite{deshpande,breaking}. The minimal case is argued to be
adequate in achieving both the gauge coupling unification and breaking of
$SO(10)$ to the SM through an intermediate scale. The exact value of the
required intermediate scale depend on the chain of the \10 breaking and various
cases are given in \cite{bertolini3}. The $45_H$ field contains components
transforming as  (15,1,1) and (1,1,3) under the Pati-Salam group
$SU(4)\times SU(2)_L\times  SU(2)_R$. They allow $SO(10)$ braking to
$SU(3)_C\times SU(2)_L\times
SU(2)_R\times U(1)_{B-L}$, $SU(4)\times SU(2)_L\times U(1)_R$ or to $SU(5)\times U(1)$ groups. At
the tree level, only $SU(5)$ intermediate stage is shown to lead to consistent spectra without
tachyons \cite{breaking} but this chain does not preserve the gauge coupling unification. As
discussed in \cite{bertolini4} turning on 1-loop corrections also allows the other two breaking
chains which preserve gauge coupling unification. The final breaking to SM can be achieved by the
(1,1,3) component of $\overline{126}_H$ which also leads to neutrino masses. The $10_H$ and
$\overline{126}_H$ contain respectively bi-doublets (1,2,2) and (15,2,2). They need to mix in order
to finally generate the standard model doublet(s) simultaneously containing  the $10_H$ and
$\overline{126}_H$ components. This can be achieved by fine tuning. For example, the mixing between
bi-doublets is achieved through the following term 
\be \label{mixing}
V\sim \chi_{ij}\chi_{kl}\Sigma_{ijklm}\phi_m\ee
which couples $45_H$ ($\chi$) to $\overline{126}_H$ ($\Sigma$) and $10_H$ ($\phi$). This mixes two
bi-doublets when component of $45_H$ transforming as singlet under the $SU(3)_c\times SU(2)_L\times
SU(2)_R\times U(1)_{B-L}$ acquires a vev. Then through fine tuning one can keep one of the two
bi-doublets in $10_H$ and $\overline{126}_H$ at the intermediate scale.
Subsequent breaking to SM is achieved through the $(1,1,3)$ component of
$\overline{126}_H$. Eq.(\ref{mixing}) provides this way
the required mixing  between doublets in $\overline{126}_H$ and $10_H$.\\

\subsection{Numerical Analysis: Model with only $10+\overline{126}$ Higgs fields.}
A non-supersymmetric \10 model with $10$ and $\overline{126}$ Higgs fields
together with $U(1)_{PQ}$ symmetry has the same Yukawa interactions as the 
minimal SUSY \10, eq.(\ref{genmass}) with $G=0$.
Minimization is performed based on the input values of the charged fermion
masses obtained by running quark and lepton masses up to the GUT scale with
$m_H$=140 GeV \cite{xing}. We use the updated low energy values of quark
mixing angles, CP phase and neutrino parameters since the effect of RG
is known to be negligible for hierarchical neutrino spectrum. We reproduce all
the input values in Table(\ref{tab:nonsusyinput}) for convenience. As before, we
take $M_d$ and $M_l$ as independent and express the remaining matrices in terms
of them and $r,s$ as in eq.(\ref{genmassmin}). Since the masses of the charged
leptons are known precisely, we go to the basis with a diagonal $M_l$ and use
them as fixed input. Thus we have 15 real parameters (12 in $M_d$, complex $s$
and real $r$) which determine remaining 13 observables shown in
Table~(\ref{tab:nonsusyinput}). The $\chi^2$ function is defined in terms of
these parameters.\\
\begin{small}
\begin{table} [ht]
\begin{math}
\begin{array}{|c|c||c|c|}
\hline
 \multicolumn{4}{|c|}{\text{GUT scale values
 with propagated uncertainty} }\\
\hline
 m_d(\text{MeV}) &1.14^{+0.51}_{-0.48}  	&\dms(\text{eV$^2$})&(7.59\pm 0.20)\times10^{-5} \\ 
 m_s(\text{MeV}) &22^{+7}_{-6} 		&\Dma(\text{eV$^2$})&(2.51\pm 0.12)\times
10^{-3}\\
 m_b(\text{GeV}) & 1.00\pm0.04			& \sin  \theta _{12}^q & 0.2246\pm 0.0011 \\
 m_u(\text{MeV}) & 0.48^{+0.20}_{-0.17} 	& \sin  \theta _{23}^q & 0.0420\pm 0.0013 \\
 m_c(\text{GeV}) & 0.235^{+0.035}_{-0.034} 	& \sin  \theta _{13}^q & 0.0035\pm 0.0003 \\
 m_t(\text{GeV}) &74.0^{+4.0}_{-3.7}		& \sin ^2 \theta _{12}^l & 0.3208\pm 0.0164 \\
 m_e(\text{MeV}) &0.469652046\pm0.000000041 	&\sin ^2 \theta _{23}^l&0.4529 ^{+0.0924}_{-0.0484}
\\
 m_{\mu }(\text{MeV})&99.1466226\pm0.0000089 	& \sin ^2\theta _{13}^l & < 0.049 (3\sigma) \\
 m_{\tau }(\text{GeV})&1.68558\pm0.00019	&\delta_{CKM}&69.63^{\circ }\pm 3.3^{\circ }\\
\hline
\end{array}
\end{math}
\vspace{0.5cm}
\caption{Input values for quark and leptonic masses and mixing angles in the non-supersymmetric
standard model extrapolated  at $M_{GUT}=2\times 10^{16}$ GeV.}
\label{tab:nonsusyinput}
\end{table}
\end{small}

Results of numerical analysis carried out separately for type-I and type-II
dominated seesaw mechanisms are shown in Table(\ref{nsmin}). Parameters obtained
for the best fit solutions are shown in Appendix A. It is evident that the
type-II mechanism fails completely in reproducing the spectrum. Once again this
is linked to the complete absence of the $b$-$\tau$ unification in
non-supersymmetric theories. Neither the atmospheric mixing nor the $b$ quark
mass can be reproduced correctly in this fit. In contrast, the type-I seesaw
works quite well. In fact, the quality of fit in this case is much better than
the minimal supersymmetric model with type-I seesaw, Table~(\ref{smint1}).
\begin{table} [ht]
\begin{small}
\begin{math}
\begin{array}{|c||c|c||c|c|}
\hline
  & \multicolumn{2}{|c||} {\text{Type-I}} & \multicolumn{2}{|c|} {\text{Type-II}} \\
\hline
 \text{Observables} & \text{Fitted value} & \text{pull} & \text{Fitted value} &
\text{pull} \\
\hline
 m_d & 0.000810163 & -0.687161 & 0.00101285 & -0.264898 \\
 m_s & 0.0208099 & -0.198354 & 0.0225915 & 0.0844982 \\
 m_b & 0.999667 & -0.00831657 & 1.08201 & 2.05031 \\
 m_u & 0.000495023 & 0.0751133 & 0.000507336 & 0.13668 \\
 m_c & 0.237348 & 0.0670883 & 0.237096 & 0.0598882 \\
 m_t & 73.9427 & -0.0154941 & 74.3006 & 0.075144 \\
 m_e & 0.000469652 & - & 0.000469652 & - \\
 m_{\mu } & 0.0991466 & - & 0.0991466 & - \\
 m_{\tau } & 1.68558 & - & 1.68558 & - \\
 \left( \dfrac{\dms}{\Dma}\right) & 0.030526 & 0.127968 & 0.0297114 & -0.235285 \\
 \sin  \theta _{12}^{q} & 0.224651 & 0.0464044 & 0.224499 & -0.0916848 \\
 \sin  \theta _{23}^{q} & 0.0420499 & 0.0392946 & 0.0421308 & 0.103004 \\
 \sin  \theta _{13}^{q} & 0.00349369 & -0.0974312 & 0.00353053 & 0.0389979 \\
 \sin ^2 \theta _{12}^{l} & 0.323245 & 0.148134 & 0.3108 & -0.610792 \\
 \sin ^2 \theta _{23}^{l} & 0.435096 & -0.369178 & 0.113306 & -7.02461 \\
 \sin ^2 \theta _{13}^{l} & \textbf{0.0244287} & - & \textbf{0.0176863} &- \\
 \delta _{\text{CKM}}[^{\circ}] & 69.5262 & -0.0314447 & 69.2051 & -0.128759 \\
 \delta _{\text{MNS}}[^{\circ}] & \textbf{318.465} & - & \textbf{14.5386} & - \\
 \alpha _1[^{\circ}] & \textbf{21.5053} & - & \textbf{345.645} & - \\
 \alpha _2[^{\circ}] & \textbf{215.128} & - & \textbf{141.905} & - \\
 r_{R(L)} & \textbf{5.62}\times \textbf {10}^{\textbf{-14}} & - &
\textbf{2.09}\times\textbf{10}^{\textbf{-10}} & - \\
 \hline
\hline
 \chi^2 &    & \textbf{0.710777} &    & \textbf{54.1197}\\
\hline
\end{array}
\end{math}
\end{small}
\vspace{0.0cm}
\caption{Best fit solutions for fermion masses and  mixing obtained assuming the type-I and type-II
seesaw dominance  in the minimal non-SUSY \10 model. Various observables and their pulls at the
minimum are shown. All the masses shown are in GeV units. The bold faced
quantities are predictions of the respective solutions.}
\label{nsmin}
\end{table}

As before the $r_R$ gets determined from the atmospheric neutrino mass scale.
Assuming, that only
one standard model survives at the electroweak scale one has,
$$\vev{(1,3,-2)}_{\overline{126}_H} \approx r_R^{-1} s_m v$$
$r_R$ in Table~(\ref{nsmin}) gives 
\be \label{scale2}\vev{(1,3,-2)}_{\overline{126}_H } \approx 3\times 10^{15} s_m  {\rm GeV} ~.\ee
Unlike the supersymmetric model, one would like to have this scale at an intermediate value
$10^{11}$ GeV \cite{bertolini3} in order to achieve the gauge coupling unification. This will
require substantial fine tuning. The exact value of the required intermediate scale for the gauge
coupling unification would depend on threshold effects not included in
the analysis in \cite{bertolini3}. This would need a detailed study of the scalar potential
minimization and the scalar sector of the theory.\\

\begin{figure}[ht]
 \centering
 \includegraphics[width=8cm,height=5cm]{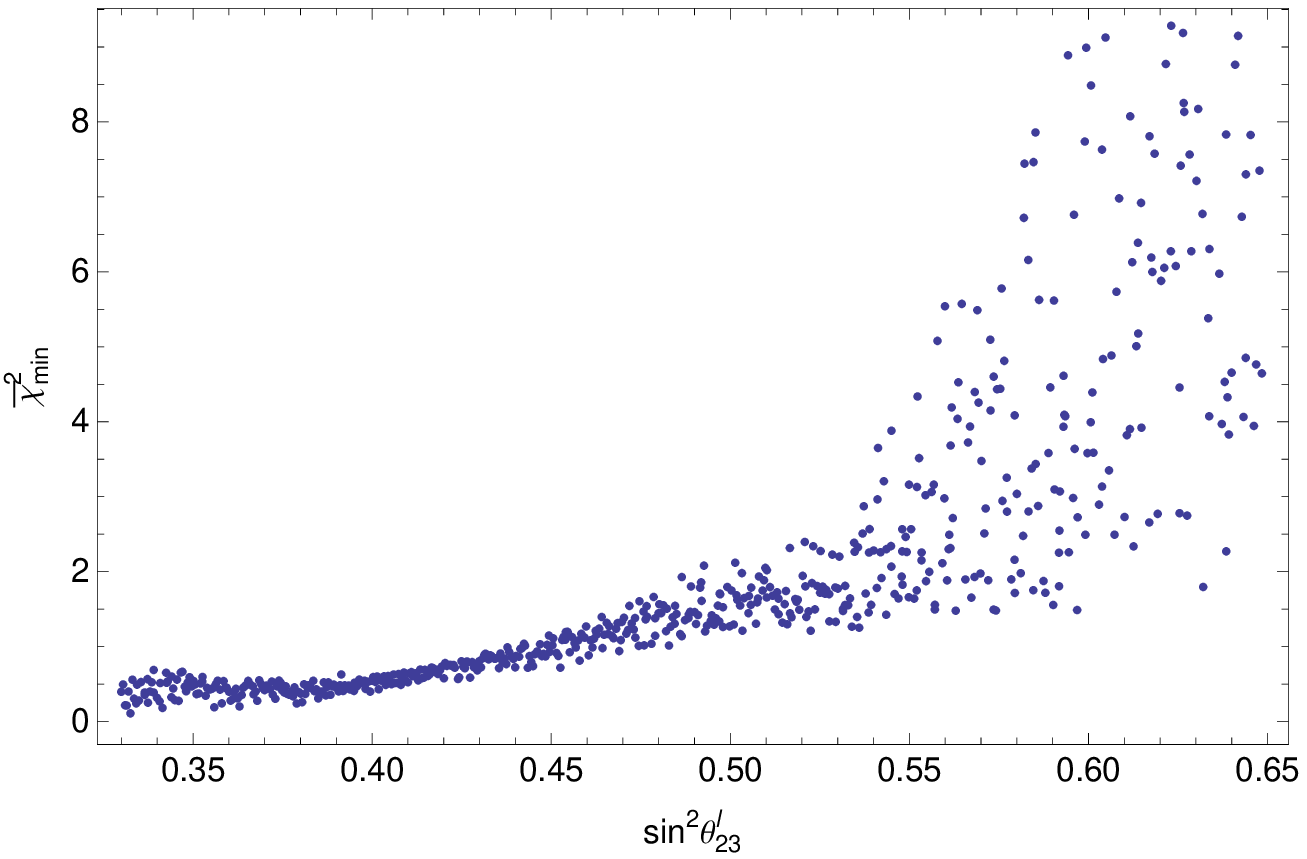}
 \caption{Variation of $\bar{\chi}^2_{min}$ with $\sin^2\theta_{23}^{l}$ in the
minimal non-susy \10 model with Type-I seesaw.}
 \label{fig1}
\end{figure}

\begin{figure}[ht]
 \centering
 \includegraphics[width=8cm,height=5cm]{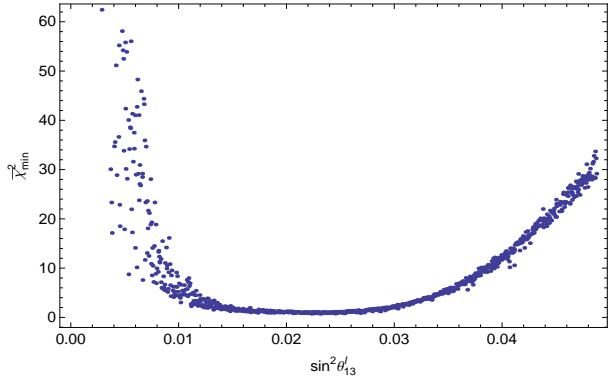}
 \caption{Variation of $\bar{\chi}^2_{min}$ with $\sin^2\theta_{13}^{l}$ in the
minimal non-susy \10
model with Type-I seesaw.}
 \label{fig2}
\end{figure}

The leptonic parameters $\theta_{13}$ and three CP violating phases
$\alpha_{1,2}$ and
$\delta_{MNS}$ get fixed at the minimum and are shown in Table(\ref{nsmin}). The
firm predictions on these observables in the scheme can be obtained by checking
the variation of $\chi^2$ with the values of various observables. Following,
\cite{grimus2,mutau}, we pin down a specific value $p_0$ of an observable $P$ by
adding a term $$\chi_P^2=\left(\frac{P-p_0}{0.01~p_0}\right)^2$$ to $\chi^2$
and then minimizing $$\hat{\chi}^2\equiv \chi^2+\chi_P^2~.$$ If $P$ happens to
be one of the observables used in defining $\chi^2$, then its contribution is
removed from there. Artificially introduced small error fixes the value $p_0$ 
for $P$ at the minimum of the $\hat{\chi}^2$. We then look at the variation of  
\be \label{chibar}
\bar{\chi}^2_{min}\equiv (\hat{\chi}^2 - \chi_P^2) |_{min}
\ee 
with $p_0$. The results of such analysis carried out for the observables $\sin^2\theta_{23}^l$ and
$\sin^2\theta_{13}^l$ are displayed in Fig.\ref{fig1} and Fig.\ref{fig2} respectively.
$\sin^2\theta_{23}^l$ can assume value in large range and the 90\% confidence level bound
corresponding to $\Delta\chi^2=4.61$ covers its entire $3\sigma$ range $0.33-0.64$. In contrast, a
clear prediction emerges for the angle $\sin^2\theta_{13}^l$ which preferentially lies in
the range $0.015-0.03$.

\subsection{Numerical Analysis: Model with only $120+\overline{126}$ Higgs fields.}
We now consider an alternative model obtained by replacing $10_H$ with $120_H$ in the minimal model
discussed before. This model is argued to be quite attractive and predictive when restricted to the
second and the third generations \cite{goran}. It is thus interesting to see if the model works in
more realistic case with three generations which require explanation of several
new parameters.\\

The fermion mass relations in this model are given by eq.(\ref{genmass}) with
$H=0$. $F$ can be made real diagonal without loss of generality. $M_l$ is not
diagonal and the charged lepton masses are
included in the $\chi^2$ function (\ref{chisq}) unlike the previous case of the
minimal model where
they were set as input. Since the errors in the charged lepton masses are extremely small, the
numerical optimization algorithm we use is unable to converge to the solution in finite time. Thus
we set 10\% error in charged lepton masses and minimize the $\chi^2$ with respect to  16 (3 in
$F$, 6 in $G$, real $s$ and complex $t_l,t_u,t_D$) real parameters.
\begin{table} [ht]
\begin{small}
\begin{math}
\begin{array}{|c||c|c||c|c|}
\hline
  & \multicolumn{2}{|c||} {\text{Type-I}} & \multicolumn{2}{|c|} {\text{Type-II}} \\
\hline
 \text{Observables} & \text{Fitted value} & \text{pull} & \text{Fitted value} &
\text{pull} \\
\hline
 m_d & 0.000186192 & -1.9871 & 0.000223284 & -1.90982 \\
 m_s & 0.00267758 & -3.2204 & 0.00296063 & -3.17323 \\
 m_b & 0.844022 & -3.89946 & 0.836471 & -4.08822 \\
 m_u & 0.00048096 & 0.00480131 & 0.000483412 & 0.0170595 \\
 m_c & 0.23454 & -0.0135291 & 0.237869 & 0.0819818 \\
 m_t & 74.053 & 0.0132566 & 73.891 & -0.0294532 \\
 m_e & 0.000467656 & -0.0424983 & 0.000475465 & 0.123771 \\
 m_{\mu } & 0.0964545 & -0.271534 & 0.101839 & 0.271587 \\
 m_{\tau } & 2.61149 & 5.49314 & 2.60147 & 5.43367 \\
 \left( \dfrac{\dms}{\Dma}\right) & 0.0303749 & 0.0605819 & 8.59\times10^{-7} & -13.4841 \\
 \sin  \theta _{12}^{q} & 0.224581 & -0.0172464 & 0.224591 & -0.00790894 \\
 \sin  \theta _{23}^{q} & 0.0419722 & -0.0218756 & 0.0420623 & 0.0490417 \\
 \sin  \theta _{13}^{q} & 0.00354561 & 0.0948516 & 0.00353062 & 0.0393252 \\
 \sin ^2 \theta _{12}^{l} & 0.321216 & 0.0243762 & 0.320612 & -0.0124452 \\
 \sin ^2 \theta _{23}^{l} & 0.450311 & -0.0544896 & 0.0375094 & -8.59228 \\
 \delta _{\text{CKM}} & 69.5526 & -0.0234639 & 69.5794 & -0.0153481 \\
 \hline
\hline
 \chi^2 &    & \textbf{59.7934} &    & \textbf{315.705}\\
\hline
\end{array}
\end{math}
\end{small}
\vspace{0.0cm}
\caption{Best fit solutions for fermion masses and  mixing obtained assuming the type-I and type-II
seesaw dominance in non-SUSY \10 model with $120+\overline{126}$ Higgs. All the
masses shown are in GeV units. Various observables
and their pulls at the minimum are shown.}
\label{ns120}
\end{table}

Results of numerical analysis carried out separately for type-I and type-II
dominated seesaw
scenarios are shown in Table(\ref{ns120}). The detailed fits are quite different
in two cases showing that a simple proportionality of the type-II and type-I
contribution observed in the two generation study \cite{goran} does not hold in
general. The model fails badly in reproducing the fermion mass spectrum in
either case. Analytic study of the two generations lead in the model to a
relation $m_{\tau} \approx 3 m_b$. This is born out in the detailed numerical
study with three generations as well. But this relation becomes one of the
causes of the failure of the model as is clearly seen in the Table
(\ref{ns120}). Likewise, the numerical fits lead to nearly vanishing solar scale
at the minimum in the type-II case. This becomes an added cause of very poor
fits. It appears from the results that the renormalizable model with
$45+120+\overline{126}$ Higgs fields is not a good candidate to obtain even 
fermion mass spectrum in spite of its attractiveness at the two generation
level \cite{goran}.\\

\subsection{Numerical Analysis: Model with $10+\overline{126}+120$ Higgs fields.}
As before, we consider the case with Hermitian (Dirac) mass matrices. The mass
relations are same as in eq.(\ref{genmass}) with all parameters real. We have
chosen the basis with a diagonal $H$. $M_l$ is not diagonal in this basis and we
parameterize it as $M_l=U_lD_lU_l^\dagger$ with $U_l$ being a general unitary
matrix expressed in terms of three angles and six phases and $D_l$ is a
diagonal matrix for the charged lepton masses. One can rewrite $M_l$ in
eq.(\ref{genmass}) as
$$3 F-it_lG=H-U_lD_lU_l^\dagger$$ 
Since $F$ and $G$ are real, the real and imaginary parts of the RHS separately determine $F$ and
$t_l G$ in terms of the charged lepton masses and parameters of $H$ and $U_l$ which are put back in
eq.(\ref{genmass}). The remaining fermion mass matrices can be expressed in
terms of 17 (3 in $H$, 9 in $U_l$, real $r,s,t_l,t_u,t_D$) real parameters in
the case of type-I seesaw dominance which
determine 16 observables $P_i$ shown in Table(\ref{tab:nonsusyinput}). One
parameter $t_D$ becomes irrelevant for the type-II seesaw case. We do the
numerical analysis for this case and results are shown in Table(\ref{nsscpv}).

\begin{table} [ht]
\begin{small}
\begin{math}
\begin{array}{|c||c|c||c|c|}
\hline
  & \multicolumn{2}{|c||} {\text{Type-I}} & \multicolumn{2}{|c|} {\text{Type-II}} \\
\hline
 \text{Observables} & \text{Fitted value} & \text{pull} & \text{Fitted value} &
\text{pull} \\
\hline
 m_d & 0.00113968 & -0.000676838 & 0.00108711 & -0.110189 \\
 m_s & 0.0219909 & -0.00150966 & 0.0142689 & -1.28852 \\
 m_b & 1. & 0.0000376219 & 1.19665 & 4.9162 \\
 m_u & 0.000480133 & 0.000666686 & 0.000486627 & 0.0331338 \\
 m_c & 0.235007 & 0.000211758 & 0.240819 & 0.166268 \\
 m_t & 73.9997 & -0.0000888053 & 77.4295 & 0.857367 \\
 m_e & 0.000469652 & 0 & 0.000469652 & 0 \\
 m_{\mu } & 0.0991466 & 0 & 0.0991466 & 0.220249 \\
 m_{\tau } & 1.68558 & 0 & 1.68558 & 0.000124065 \\
 \left( \dfrac{\dms}{\Dma}\right) & 0.0302402 & 0.000545016 & 0.0260106 & -1.88556 \\
 \sin  \theta _{12}^{q} & 0.224601 & 0.00105776 & 0.224567 & -0.0304356 \\
 \sin  \theta _{23}^{q} & 0.0420001 & 0.0000431604 & 0.0431393 & 0.897068 \\
 \sin  \theta _{13}^{q} & 0.00351992 & -0.000308192 & 0.00338234 & -0.509862 \\
 \sin ^2 \theta _{12}^{l} & 0.320821 & 0.000292661 & 0.278093 & -2.6052 \\
 \sin ^2 \theta _{23}^{l} & 0.453034 & 0.000947066 & 0.343286 & -2.26804 \\
 \sin ^2 \theta _{13}^{l} & \textbf{0.0306736} & - & \textbf{0.00538748} & - \\
 \delta _{\text{CKM}}[^{\circ}] & 69.6278 & -0.000660788 & 72.7155 & 0.935014 \\
 \delta _{\text{MNS}}[^{\circ}] & \textbf{355.719} & - & \textbf{46.8148} & - \\
 \alpha _1[^{\circ}] & \textbf{60.079 }& - & \textbf{60.6202} & - \\
 \alpha _2[^{\circ}] & \textbf{214.691} & - & \textbf{250.978} & - \\
 r_{R(L)} & \textbf{1.56}\times\textbf{10}^{\textbf{-15}} & - &
\textbf{3.43}\times\textbf{10}^{\textbf{-10}} & - \\
 \hline
\hline
 \chi^2 &   & \sim \textbf{10}^{\textbf{-6}} &    & \textbf{44.0801}\\
\hline
\end{array}
\end{math}
\end{small}
\vspace{0.0cm}
\caption{Best fit solutions for fermion masses and  mixing obtained assuming the type-I and type-II
seesaw dominance in the non-supersymmetric  \10 model with $10+\overline{126}+120$ Higgs. Various
observables and their pulls at the minimum are shown. All the masses shown are
in GeV units. The bold faced quantities are predictions of the respective
solutions.}
\label{nsscpv}
\end{table}

Parameters obtained for the best fit solutions in type-I case are
shown in Appendix B. Unlike the supersymmetric case, the presence of $120_H$
does not help in improving the fits in the type-II seesaw dominated case. But
the fits obtained for the type-I scenario are considerably better compared to
the corresponding supersymmetric as well as the minimal non-supersymmetric case.
Pulls in all observables are practically zero in this case.\\

\begin{figure}[ht]
 \centering
 \includegraphics[width=8cm,height=5cm]{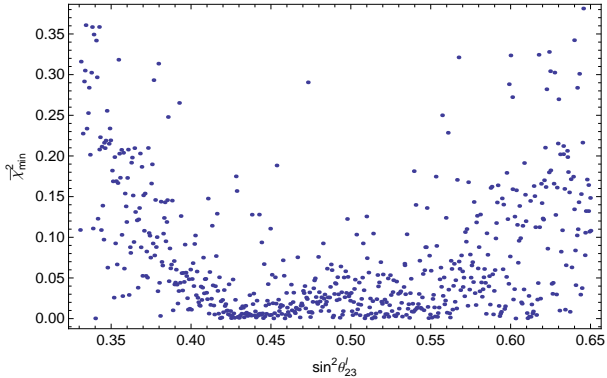}
 \caption{Variation of $\bar{\chi}^2_{min}$ with $\sin^2\theta_{23}^{l}$ in the extended model
with Type-I seesaw.}
 \label{fig3}
\end{figure}

\begin{figure}[ht]
 \centering
 \includegraphics[width=8cm,height=5cm]{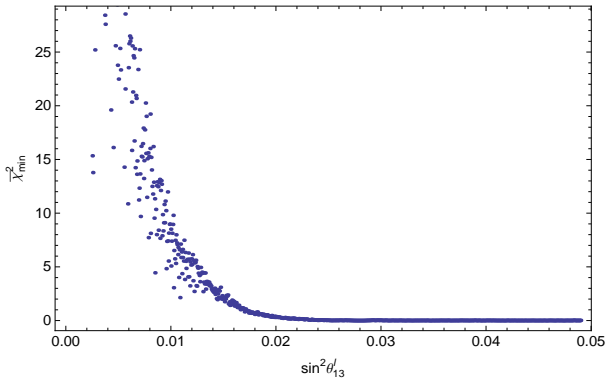}
 \caption{Variation of $\bar{\chi}^2_{min}$ with $\sin^2\theta_{13}^{l}$ in the extended model
with Type-I seesaw.}
 \label{fig4}
\end{figure}

The predictions of the model for the observables $\sin^2\theta_{23}^l$ and
$\sin^2\theta_{13}^l$ are
displayed in Fig.\ref{fig3} and Fig.\ref{fig4} respectively. Once again a clear
prediction$\sin^2\theta_{13}\gtrsim 0.015$ emerges in this case.

\section{Numerical analysis and underlying flavour structure}
Numerical analysis presented in the previous section has demonstrated viability of various \10
models in explaining the fermion masses and mixing. In the process, it has also provided us with
specific structure of fermion mass matrices which can be used to obtain some insight into the
underlying flavour structure. We discuss one specific case namely, the  minimal non-supersymmetric
model from this point of view.\\

At the \10 level, Yukawa couplings $H,F,G$ determine the flavour structure of
various mass matrices. Thus any underlying flavour symmetry if it exists should
get reflected in the structure of these matrices. Specific structures  for the
Yukawa coupling matrices have been used to predict relations
between the (hierarchical) quark masses and (small) quark mixing, see for
example \cite{ross1, ross2, ds}. In a large class of such models, the observed
masses and mixing patterns among quarks are reproduced when elements of the
quark mass matrices are expressed as powers of one or two
expansion parameters. Following this, we try to look for a  similar
parameterization for the underlying matrices $F,H$ in case of the minimal
non-supersymmetric model. We choose the Cabibbo angle $\lambda=0.2246$ as a
convenient parameter. Elements of  $F$ and $H$ in this case are then
found to have the following hierarchical structure in the basis with a diagonal
$M_l$:

\begin{footnotesize}\beqa \label{HF}
H&=1.088 e^{0.435 i} {\rm ~GeV}&\left(
\begin{array}{ccc}
 0.513 e^{1.659i} \lambda^4&0.361 e^{-1.257i}\lambda^3 &0.685 e^{0.843i}\lambda^2 \\
 0.361 e^{-1.257i}\lambda^3& 0.119e^{1.143i}\lambda^2&0.490 e^{-2.123i}\lambda \\
0.685 e^{0.843i}\lambda^2& 0.490 e^{-2.123i}\lambda& 1
\end{array}
\right)~,\nonumber \\
F&=0.278 e^{2.561 i}{\rm ~ GeV}&\left(
\begin{array}{ccc}
 0.802 e^{- 0.226i} \lambda^4&0.470 e^{2.90i}\lambda^3 &0.892 e^{-1.283i}\lambda^2 \\
0.470 e^{2.90i}\lambda^3& 2.359e^{0.515i}\lambda^2&0.639e^{2.034i}\lambda \\
0.892 e^{-1.283i}\lambda^2& 0.639e^{2.034i}\lambda & 1
\end{array}
\right)  \eeqa \end{footnotesize}
$33$ element turns out to be largest both for $F$ and $H$ and we have normalized other elements by
its value in writing the above structure. Most coefficients in powers of $\lambda$ are roughly ${\cal
O}(1)$ except for the $22$ elements.\\

The above structure determined numerically here is suggestive of an underlying $U(1)$ symmetry used
\cite{fg} in the Froggatt Nielsen (FG) approach. Indeed a simple $U(1)$ can explain the occurrence
of various powers of $\lambda$ in eq.(\ref{HF}). Consider a $U(1)$ symmetry with the $U(1)$ charges
$2,1,0$ assigned respectively to three generations of $16_F$-plet. Both $10_H$ and
$\overline{126}_H$ are assumed neutral under this symmetry. In this case, the $33$ elements of $F,H$
arise from the renormalizable couplings $16_{3F}16_{3F}\phi_H~~~(\phi=10,\overline{126} )$. The $23$
and $32$ elements follow from the couplings $16_{2F}16_{3F}\phi_H\frac{\eta}{M}$. Likewise, the two,
three and four powers of $\eta$ respectively  generate ${\cal O}(\lambda^2,\lambda^3,\lambda^4)$
terms in eq.(\ref{HF}) where $\lambda=\frac{\vev{\eta}}{M}$, $M$ being some underlying scale above
the $U(1)$ breaking scale $\vev{\eta}$ and $\eta$ is assumed to carry the $U(1)$ charge $-1$. The
quark mass matrices resulting from the above $F,H$ also follow this simple pattern as
in eq.(\ref{HF}):
\begin{footnotesize}\beqa \label{mumd}
M_d&=0.9708 e^{0.6809 i}{\rm ~ GeV}&\left(
\begin{array}{ccc}
 0.800 e^{1.481i} \lambda^4&0.539 e^{-1.503i}\lambda^3 &1.024 e^{0.597i}\lambda^2 \\
 0.539 e^{-1.503i}\lambda^3& 0.699e^{2.204i}\lambda^2&0.733 e^{-2.370i}\lambda \\
1.024 e^{0.597i} \lambda^2& 0.733 e^{-2.370i}\lambda& 1
\end{array}
\right)~,\nonumber \\
M_u&=72.639 e^{0.523 i}{\rm ~ GeV}&\left(
\begin{array}{ccc}
 0.609 e^{1.585i} \lambda^4&0.419 e^{-1.359i}\lambda^3 &0.796 e^{0.741i}\lambda^2 \\
 0.419 e^{-1.359i}\lambda^3& 0.281e^{1.981i}\lambda^2&0.570e^{-2.225i}\lambda \\
 0.796 e^{0.741i}\lambda^2& 0.570e^{-2.225i}\lambda& 1
\end{array}
\right) \nonumber \eeqa \end{footnotesize}

This structure is already proposed and studied  in \cite{ds} as a possible explanation of 
quark and neutrino mixing and masses. Here it follows from a detailed analysis of this specific \10
model. As shown in \cite{ds}, such a form can reproduce the observed mixing and mass patterns for
quarks. The expansion parameter chosen in \cite{ds} is somewhat larger, $\lambda=0.26$. The Dirac
neutrino mass matrix on the other hand is given by
\begin{footnotesize}\be 
M_D = 86.240 e^{0.210 i} {\rm ~ GeV}\left(
\ba{ccc}
 0.253 e^{1.795 i} \lambda ^4 & 0.201 e^{-0.959 i} \lambda ^3 & 0.382 e^{1.140 i}
\lambda ^2 \\
 0.201 e^{-0.959 i} \lambda ^3 & 0.567 e^{-0.222 i} \lambda ^2 & 0.273 e^{-1.825 i}
\lambda  \\
 0.382 e^{1.140 i} \lambda ^2 & 0.273 e^{-1.825 i} \lambda  & 1
\ea \right) \ee       \end{footnotesize}

The coefficients in front of various elements are anomalously small and thus $M_D$ does not really
share the same symmetry as the underlying Yukawa matrices. The $M_D$ and $M_R\sim F$ conspire to
produce a neutrino mass matrix which has an interesting form
\begin{footnotesize}\be 
M_\nu=0.087 e^{-0.898 i} r_R r^2 {\rm ~ GeV}\left(
\ba{ccc}
 1.339 e^{2.543 i} \lambda ^3 & 0.878126 e^{-0.662 i} \lambda ^2 & 1.753 e^{1.529 i}
\lambda  \\
 0.878 e^{-0.662 i} \lambda ^2 & 0.800 e^{-2.646 i} & 1.062 e^{-1.458 i} \\
 1.753 e^{1.529 i} \lambda  & 1.062 e^{-1.458 i} & 1
\ea \right) \ee       \end{footnotesize}

Since we are working in a basis with a diagonal $M_l$, the above matrix
determines physical neutrino mixing and  allows us to understand the leptonic
mixing structure analytically. Firstly, the $23$ block has all elements of
${\cal O}(1)$ which results in the large atmospheric angle and hierarchy
in neutrino masses. Secondly, the $11$ and $12$ elements are zero to leading
order in $\lambda$. In the approximation of neglecting higher powers of
$\lambda$, the $M_\nu$ has two-zero texture
(classified as A1 in \cite{glashow}). The presence of the zeros leads to a firm
prediction of the
third mixing angle \cite{glashow}
\be \label{zeros}
\sin^2\theta_{13}^l\approx \left(\frac{\Delta m_{sol}^2}{ \Delta m_{atm}^2}\right)\frac{\sin^2 \theta_{12}^l \cos^2 \theta_{12}^l}{\cos2\theta_{12}^l\tan^2\theta_{23}^l}~. \ee
This analytic relation is in very good agreement with the numerical values. Evaluation of the RHS
using the best fit values of parameters in Table(\ref{nsmin} ) leads to $\sin^2\theta_{13}^l\approx
0.0245$ in agreement with the numerical prediction. Even away from the minimum $\chi^2$, one would
get $\sin^2\theta_{13}^l$ around $ 0.02$ as long as two zero structure and
hence eq.(\ref{zeros}) holds approximately. This is born out quite well in
Figure \ref{fig2}.\\

The simple $U(1)$ symmetry used to explain the structure of $F,H$ may appear to
have two shortcomings. Firstly, the specific structures are found in a basis
with a diagonal $M_l$ Secondly, the coefficients of powers of $\lambda$ in $F,H$
are not strictly ${\cal O}(1)$, notably in the $22$ elements. In general, the
definition of symmetry and  resulting texture of Yukawa matrices are basis
dependent. Basis with a diagonal $M_l$ are very special basis and it would be
more desirable to find a basis in which $M_l$ also has a structure similar to
the $F,H,M_d,M_u$. One can indeed find a class of unitary rotations which bring
the diagonal $M_l$ to the form as in eq.(\ref{HF}) and at the same
time retain the forms of $F,H$ albeit with a different set of coefficients. The
$U(1)$ symmetry leads to the following general form of the Yukawa matrices:
\be \label{form}
(F,H)=a_{33}^{F,H} \left(\ba{ccc}
a_{11}^{F,H}\lambda^4&a_{12}^{F,H}\lambda^3&a_{13}^{F,H}\lambda^2\\
a_{12}^{F,H}\lambda^3&a_{22}^{F,H}\lambda^2&a_{23}^{F,H}\lambda\\
a_{13}^{F,H}\lambda^2&a_{23}^{F,H}\lambda& 1 \ea\right)~. \ee
$F,H$ as given above can be diagonalized with high accuracy by rotation $R_{F,H}$ consisting of three successive rotations in $2-3$, $1-3$ and $1-2$ plane with the
 mixing angles \cite{ds}
\beqa \label{angles} 
\sin \theta_{23}^{F,H} &\approx&  a_{23}^{F,H}~\lambda, \nonumber \\
\sin \theta_{13}^{F,H} &\approx&  a_{13}^{F,H}~\lambda^2, \nonumber \\
\tan 2\theta_{12}^{F,H} &\approx&  2 \lambda~ \dfrac{a_{12}^{F,H}-a_{23}^{F,H}
a_{13}^{F,H}}{a_{22}^{F,H}-(a_{23}^{F,H})^2+{\cal O}(\lambda^2)} \eeqa

The eigenvalues of $F,H$ are ${\cal O}(1,\lambda^2,\lambda^4)$. The eigenvalues of $M_l$
are roughly of similar order-though the coefficient for the electron mass is
somewhat small. Thus $M_l$ can be put to the form as in (\ref{HF}) by rotating
the diagonal $M_l$ with a rotation matrix $V_l$ with  angles as in
eq.(\ref{angles}) but with a different set of coefficients $a_{ij}^l$.  It is
easy to see that when $F,H$ are expressed in new basis their forms do not change
to leading order in $\lambda$ but now  coefficients in front of powers of
$\lambda$ are different say, $a_{ij}'^{F,H}$. They depend on $a_{ij}^{F,H}$ and
$a_{ij}^l$. Thus symmetry in question may manifest itself in more general basis
than the specific diagonal basis provided  $a_{ij}'^{F,H}$ are also ${\cal
O}$(1).\\

Let us consider a simple example. Rotate $F,H$ and diagonal $M_l$ with a common
rotation $V_l$ defined as 
$$V_l=\left(\ba{ccc}1&0&0\\
0&1-1/2 \lambda^2&\lambda e^{i \beta}\\
0& -\lambda e^{-i \beta}&1-1/2 \lambda^2\\
\ea\right)$$
$\beta$ can be chosen such that the coefficient of various powers of $\lambda$ in elements of
$F'=V_l^T FV_l$ and $H'=V_l HV_l$ are near to 1. The best fit value of $\beta$ turns out to be
$\beta=1.055$ and for this one gets
\begin{footnotesize}
\beqa
|F'|&=&0.278~ {\rm GeV}\left(
\begin{array}{ccc}
0.802\lambda^4 &0.772 \lambda^3&0.892\lambda^2 \\
0.772\lambda^3 &1.12 \lambda^2 &1.638 \lambda\\
0.892\lambda^2& 1.638\lambda& 1\\
\end{array}
\right) \nonumber \\
|H'|&=&1.088~{\rm GeV}\left(
\begin{array}{ccc}
0.513\lambda^4 &0.593 \lambda^3&0.685\lambda^2 \\
0.593\lambda^3 &0.939\lambda^2 &0.876 \lambda\\
0.685\lambda^2&0.876\lambda& 1\\
\end{array}
\right) \nonumber \\     
\eeqa
\end{footnotesize}
Unlike in eq.(\ref{HF}), all the cofficients of various elementsin the above equation are now ${\cal O}(1)$. 
$M_l$ is non-diagonal in this basis and is given by
\be |M_l|=1.685 {\rm GeV}\left(
\begin{array}{ccc}
0.109\lambda^4 &0&0\\
0&1.06\lambda^2 &1.006 \lambda\\
0&1.006\lambda& 1\\
\end{array}
\right) ~.\ee

We note that the Yukawa coupling matrices in cases other than the minimal also display 
hierarchical structure, see results in Appendix (B). Thus these cases can also be understood in
terms of some simple pattern as in the minimal case discussed here explicitly.\\

\section{Summary} \10 models have been used to obtain a unified description of fermion masses and
mixing angles. We have undertaken in this paper an exhaustive analysis of many different \10 models.
Using several different data sets as input, we have numerically determined viability of these models
in reproducing the fermion spectrum. In case of the supersymmetric models, we used data
corresponding to different values of $\tan\beta$ and with or without appreciable finite
threshold correction. Comparison of different set clearly brings out an important feature. In the
minimal model with type-II seesaw dominance, the  $b$-$\tau$ unification appears to be a key
ingredient. The cases without such unification cannot explain the entire fermion spectrum. In
particular, the case of very low $\tan\beta$ showing this unification works much better than the
previously studied data set with $\tan\beta=10$. This connection is not required if neutrinos
obtain their masses from the type-I seesaw mechanism. In this case one can obtain very good fits in
the minimal model almost for every data set used, see Tabel(III). Moreover, the $B-L$ breaking scale
inferred from neutrino masses also lies closer to the GUT scale compared to the type-II seesaw mechanism.
The situation becomes better when a 120-plet of Higgs field is added to
the model. Here one can get excellent  fits to fermion masses in both the type-I and type-II seesaw
mechanisms.\\

We also carried out a detailed analysis of the fermion masses in non-supersymmetric models. The
minimal non-supersymmetric model with $45_H+10_H+\overline{126}_H$ is quite economical and is argued
recently \cite{bertolini3, bertolini4} to be a viable candidate for the gauge coupling unification.
As shown here it also provides a very good description of fermion masses as well. Intermediate scale
$\sim 10^{11}$ GeV is required in this model in order to obtain the unification of gauge coupling
\cite{bertolini3}. The scale preferred from the fits to fermion masses presented here is somewhat
larger. This scale can be reduced if the admixture of the light doublet in the doublet component of
$\overline{126}_H$ is very small, see eq.(\ref{scale2}). Viability of these as well as simultaneous
analysis of the constraint from the gauge coupling unification will depend on the detailed analysis
of the scalar sector of the theory. The Yukawa coupling matrices obtained
numerically in this case display interesting structure which can be understood from a very simple
symmetry imposed at a high scale. These features coupled with its economy makes the minimal non
supersymmetric model an attractive  choice to unify basic gauge and Yukawa
interactions.\\

\noindent{\bf Acknowledgements}\\
Computations needed for the results reported in this work were done using the PRL 3TFLOP
cluster at Physical Research Laboratory, Ahmedabad. K.M.P. would like to thank Dr. Dilip K. Angom
for providing useful tips on parallel computing.\\

\section{Appendix}

We list here the fermion mass matrices using the best fit values of the parameters given in Table
\ref{nsmin} (Table \ref{nsscpv}) corresponding to the type-I seesaw mechanism in the case of minimal
(non-minimal) non-supersymmetric \10 model. All the mass matrices are expressed in GeV units.\\

\subsection{Best fit parameter values: The minimal nonsusy \10 model,
type-I seesaw mechanism (Table \ref{nsmin}).}
Parameters obtained for best fit solution.
\begin{footnotesize}
\beqa \label{prmt1min}
r&=& 69.1739;~~s=0.362941 - 0.0463175 i \nonumber\\
M_l&=&\left(
\begin{array}{ccc}
 0.000469652 & 0 & 0 \\
 0 & 0.0991466 & 0 \\
 0 & 0 & 1.68558
\end{array}
\right) \nonumber \\
M_d&=&\left(
\begin{array}{ccc}
 -0.00110182+0.00164125 i & 0.0040374-0.00434507 i & 0.0145011+0.0480084 i \\
 0.0040374-0.00434507 i & -0.0331074+0.00870484 i & -0.0187112-0.158707 i \\
 0.0145011+0.0480084 i & -0.0187112-0.158707 i & 0.754282+0.611126 i
\end{array}
\right)
 \eeqa     \end{footnotesize}

Results:
\begin{footnotesize}\beqa \label{rest1min}
M_u&=&\left(
\begin{array}{ccc}
 -0.0575896+0.0967087 i & 0.231322-0.25593 i & 0.881792+2.78041 i \\
 0.231322-0.25593 i & -0.826159+0.612181 i & -1.21531-9.21493 i \\
 0.881792+2.78041 i & -1.21531-9.21493 i & 62.9262+36.2872 i
\end{array}
\right)\nonumber \\
M_{\nu}&=&r_R r^2\left(
\begin{array}{ccc}
 -0.0000981928+0.00131563 i & 0.0000421662-0.003853 i & 0.0276444+0.0202258 i \\
 0.0000421662-0.003853 i & -0.0640659+0.0272358 i & -0.0653091-0.0653272 i \\
 0.0276444+0.0202258 i & -0.0653091-0.0653272 i & 0.054234-0.0680089 i
\end{array}
\right) \nonumber \eeqa \end{footnotesize}

\subsection{Best fit  parameter values: The non-minimal nonsusy \10 model, type-I seesaw mechanism (Table \ref{nsscpv}).}
Parameters obtained for best fit solution.
\begin{footnotesize}
\beqa \label{prmt1scpv}
r&=& -52.4173;~s= 1.61949;~t_l= 3.1751;~t_u= 0.0413014;~t_D= -11.7339. \nonumber\\
H&=&\left(
\begin{array}{ccc}
 0.00158452 & 0 & 0 \\
 0 & 0.0407501 & 0 \\
 0 & 0 & -0.330398
\end{array}
\right)
\nonumber \\
F&=&\left(
\begin{array}{ccc}
 -0.00116221 & -0.000145513 & 0.0130876 \\
 -0.000145513 & -0.0224155 & -0.00121344 \\
 0.0130876 & -0.00121344 & -0.667509
\end{array}
\right) \nonumber \\
G&=&\left(
\begin{array}{ccc}
 0 & -0.00670763 & 0.00612927 \\
 0.00670763 & 0 & -0.0437162 \\
 -0.00612927 & 0.0437162 & 0
\end{array}
\right)
 \eeqa    
\end{footnotesize}

Results:
\begin{footnotesize}\beqa \label{rest1scpv}
M_d&=&\left(
\begin{array}{ccc}
 0.00042231 & -0.000145513-0.00670763 i & 0.0130876+0.00612927 i \\
 -0.000145513+0.00670763 i & 0.0183346 & -0.00121344-0.0437162 i \\
 0.0130876-0.00612927 i & -0.00121344+0.0437162 i & -0.997907
\end{array}
\right)\nonumber \\
M_u&=&\left(
\begin{array}{ccc}
 0.0156028 & 0.0123525+0.0145214 i & -1.11099-0.0132693 i \\
 0.0123525-0.0145214 i & -0.233181 & 0.103008+0.0946413 i \\
 -1.11099+0.0132693 i & 0.103008-0.0946413 i & 73.9827
\end{array}
\right)\nonumber \\
M_l&=&\left(
\begin{array}{ccc}
 0.00507117 & 0.00043654-0.0212974 i & -0.0392628+0.019461 i \\
 0.00043654+0.0212974 i & 0.107997 & 0.00364033-0.138803 i \\
 -0.0392628-0.019461 i & 0.00364033+0.138803 i & 1.67213
\end{array}
\right)\nonumber \\
M_{\nu}&=&r_R r^2\left(
\begin{array}{ccc}
 0.243898+0.00702837 i & 0.0907917-0.0237474 i & -1.65214-0.184893 i \\
 0.0907917-0.0237474 i & 4.6052-0.0433382 i & -5.76376+0.720305 i \\
 -1.65214-0.184893 i & -5.76376+0.720305 i & 5.33677+1.11414 i
\end{array}
\right) \eeqa \end{footnotesize}

\end{document}